\documentstyle[12pt]{article}
 
\makeatletter
 
\def\section{\@startsection {section}{1}{\z@}{+3.0ex plus +1ex minus
  +.2ex}{2.3ex plus .2ex}{\normalsize\bf}}
\def\subsection{\@startsection{subsection}{2}{\z@}{+2.5ex plus +1ex
minus +.2ex}{1.5ex plus .2ex}{\normalsize\bf}}
\def\subsubsection{\@startsection{subsubsection}{3}{\z@}{+3.25ex plus
 +1ex minus +.2ex}{1.5ex plus .2ex}{\normalsize\bf}}

\def\appendix{\setcounter{section}{0} \setcounter{subsection}{0}
              \setcounter{equation}{0}
	      \@addtoreset{equation}{section}
              \def\thesection{\Alph{section}}
              \def\theequation{\thesection\arabic{equation}}}

\newcount\@tempcntc
\def\@citex[#1]#2{\if@filesw\immediate\write\@auxout{\string\citation{#2}}\fi
  \@tempcnta\z@\@tempcntb\m@ne\def\@citea{}\@cite{\@for\@citeb:=#2\do
    {\@ifundefined
       {b@\@citeb}{\@citeo\@tempcntb\m@ne\@citea
        \def\@citea{,\penalty\@m\ }{\bf ?}\@warning
       {Citation `\@citeb' on page \thepage \space undefined}}%
    {\setbox\z@\hbox{\global\@tempcntc0\csname
b@\@citeb\endcsname\relax}%
     \ifnum\@tempcntc=\z@ \@citeo\@tempcntb\m@ne
       \@citea\def\@citea{,\penalty\@m}
       \hbox{\csname b@\@citeb\endcsname}%
     \else
      \advance\@tempcntb\@ne
      \ifnum\@tempcntb=\@tempcntc
      \else\advance\@tempcntb\m@ne\@citeo
      \@tempcnta\@tempcntc\@tempcntb\@tempcntc\fi\fi}}\@citeo}{#1}}

\def\@citeo{\ifnum\@tempcnta>\@tempcntb\else\@citea
  \def\@citea{,\penalty\@m}%
  \ifnum\@tempcnta=\@tempcntb\the\@tempcnta\else
   {\advance\@tempcnta\@ne\ifnum\@tempcnta=\@tempcntb \else
\def\@citea{-}\fi
    \advance\@tempcnta\m@ne\the\@tempcnta\@citea\the\@tempcntb}\fi\fi}

\makeatother

\expandafter\ifx\csname mathrm\endcsname\relax\def\mathrm#1{{\ri #1}}\fi

\def\nn{\nonumber}

\def\beq{\begin{equation}}
\def\eeq{\end{equation}}
\def\beqar{\begin{eqnarray}}
\def\eeqar{\end{eqnarray}}
\def\barr#1{\begin{array}{#1}}
\def\earr{\end{array}}

\def\text{\textstyle}

\def\bma{\begin{equation}}
\def\ema{\end{equation}}

\def\Ga{\Gamma}

\def\si{\sigma}

\def\theenumi{\roman{enumi}}
\def\p@enumi{\theenumi}

\def\refeq#1{\mbox{(\ref{#1})}}

\def\refse#1{\mbox{Sect.~\ref{#1}}}
\def\reffi#1{\mbox{Fig.~\ref{#1}}}
\def\refapp#1{\mbox{App.~\ref{#1}}}
\def\citere#1{\mbox{Ref.~\cite{#1}}}

\newcommand{\ri}{{\mathrm{i}}}

\renewcommand{\d}{{\mathrm{d}}}

\renewcommand{\O}{{\cal{O}}}

\newcommand{\GeV}{\unskip\,\mathswitchr{GeV}}

\def\mathswitchr#1{\relax\ifmmode{\mathrm{#1}}\else$\mathrm{#1}$\fi}

\newcommand{\Pep}{\mathswitchr e^+}
\newcommand{\Pem}{\mathswitchr e^-}

\newcommand{\PW}{\mathswitchr W}
\newcommand{\PWp}{\mathswitchr W^+}
\newcommand{\PWm}{\mathswitchr W^-}
\newcommand{\PZ}{\mathswitchr Z}

\newcommand{\Pt}{\mathswitchr t}

\def\mathswitch#1{\relax\ifmmode#1\else$#1$\fi}

\def\ie{i.e.\ }

\def\ie{i.e.\ }
\def\eg{e.g.\ }
\def\cf{cf.\ }

\renewcommand{\Re}{\mathop{\mathrm{Re}}}

\begin{document}
 
\thispagestyle{empty}
\def\thefootnote{\fnsymbol{footnote}}
\setcounter{footnote}{1}
\null
\renewcommand{\baselinestretch}{1}
\Huge
\normalsize
\mbox{} \hfill WUE-ITP-96-006\\
\mbox{} \hfill hep-ph/9605420
\vskip 1cm
\vfill
\begin{center}
{\Large \bf 
High-Energy Approximation of One-Loop Feynman Integrals
\par} \vskip 2.5em
{\large {\sc M. Roth}  and {\sc A. Denner} \\[1ex]
{\normalsize \it Institut f\"ur Theoretische Physik, Universit\"at W\"urzburg\\
Am Hubland, D-97074 W\"urzburg, Germany}
\\[2ex]
\par} \vskip 1em
\end{center} \par
\vskip 2cm 
\vfil
{\bf Abstract:} \par
We provide high-energy approximations for all  one-loop 
scalar 3- and 4-point functions and the corresponding tensor integrals 
that appear in scattering processes with four external on-shell particles.
Our expressions are valid if
all kinematical invariants are much larger than the internal and external
masses. They contain all leading-order terms of the integrals.
\par
\vfill
\noindent  WUE-ITP-96-006\par
\vskip .15mm
\noindent May 1996 \par
\null
\setcounter{page}{0}
\clearpage
\def\thefootnote{\arabic{footnote}}
\setcounter{footnote}{0}

\newcommand{\M}{{\cal M}}
\newcommand{\Se}{{\cal S}}
\newcommand{\Li}{\mbox{Li}}
\newcommand{\sign}{\si}
\newcommand{\tx}{\tilde x}
\newcommand{\im}{\ri}
\newcommand{\Ln}{L}

\section{Introduction}

The experiments at LEP and the SLC have provided a large number
of high-precision experimental data. In order to compare these data
with theoretical predictions the inclusion of one-loop radiative corrections
has been inevitable, and even the leading 
two-loop corrections had to be taken into account \cite{LEP95}. 
Although forthcoming experiments as \eg at LEP2 or a Next Linear Collider 
(NLC) will not reach such a high experimental accuracy, 
nevertheless the inclusion of one-loop radiative corrections will in
general be required in the corresponding theoretical predictions. 
In fact, as the radiative corrections typically involve logarithms of 
the energy, their relevance even grows with increasing energy.

The radiative corrections to $\Pep\Pem\to f \bar f$ ($f\ne \Pt$) are relatively
simple even off the \PZ~resonance, because
only external fermions are involved, and their masses 
can be neglected at sufficiently high energies. Once the masses of the
external particles are relevant, as \eg for top-pair production, 
or external gauge bosons are present, the
number of Feynman diagrams is considerably increased, and the
evaluation as well as  the results for the one-loop radiative corrections
are much more complicated. 
In addition, owing to the presence of different 
energy scales, numerical instabilities eventually arise in the TeV range or
beyond when the standard calculational procedures are used. These instabilities
are caused by numerical cancellations originating from the recursive
reduction of the tensor integrals and---in reactions involving longitudinal 
gauge bosons---additionally from unitarity cancellations.
Evidently, it is desirable to circumvent these problems.

As the energy scale for future experiments is large compared to the
electroweak scale, a natural approach consists in neglecting all masses
as compared to the centre-of-mass energy whenever possible.
Such an approximation has been worked out for the process $\Pep\Pem\to \PWp\PWm$
\cite{W-pair}.
This calculation has shown that the one-loop corrections indeed simplify a lot 
in a high-energy approximation, and that such an approximation can
already be useful for energies around $500\GeV$.

With purely algebraic manipulations, which can easily be performed by 
computer-algebra packages such as \eg FeynCalc \cite{FeynCalc}, 
all one-loop Feynman
amplitudes can be expressed in terms of scalar and tensor integrals.
The tensor integrals can be algebraically reduced to scalar integrals 
\cite{Pa79}.
Moreover, all scalar integrals with more than four internal propagators can be
related to scalar 4-point functions \cite{Me65}. 
Consequently, all one-loop Feynman amplitudes can be reduced to 
scalar 1-, 2-, 3-, and 4-point functions. For these functions complete analytical results 
exist \cite{tH79}.

Thus, 
the expansions of the scalar integrals for high energies 
are an essential ingredient of a high-energy approximation.
The example of $\PW$-pair
production has shown that the evaluation of these approximations requires a
substantial amount of work. In order to facilitate future calculations
in the high-energy limit, it is therefore desirable to calculate and
tabulate the high-energy expansion of these integrals.

In the literature, methods have been described to evaluate Feynman integrals
in the limit of large masses and/or momenta \cite{Ch88}. However, these
methods in general do not apply when the external particles
are on their mass shell \cite{Sm93}. In this case (IR or mass) singularities can
arise if all masses are
 neglected. For UV-singular Feynman integrals,
including all 2-point integrals, no IR or mass
singularities show up such that the corresponding high-energy
approximations are simply obtained by putting all masses to zero in the
exact results.

One can, of course, use the known exact results for the 3- and 4-point
functions to derive the approximations. 
However, the necessity to keep the masses finite in order to extract the
correct singularities renders the expansion of the exact results a tedious 
exercise.

On the other hand, the Mellin-transform technique allows 
to construct approximations in a relatively simple and direct way 
as long as  only one of the kinematical invariants gets large, as it is the 
case for 3-point functions. Moreover, it turns out that this approach can also 
be used for the scalar 4-point-function despite of the fact that 
two invariants become large in the corresponding high-energy limit.
However, the Mellin-transform technique fails for the tensor 
4-point integrals.
Therefore, we used the tensor-integral reduction algorithm to construct the
corresponding high-energy approximations.
In this way we have calculated high-energy approximations
of the basic one-loop integrals
necess\-ary for scattering processes of two on-shell particles into two
on-shell particles, \ie for the most important
processes at high energy colliders.

The paper is organized as follows:
In \refse{se:3point} we describe our evaluation of the high-energy 
approximation for the scalar and tensor 3-point functions and list the 
corresponding results.
In \refse{se:4point} the same is done for the 4-point functions.
In the appendix we summarize our conventions for the Feynman integrals.

\section{High-energy approximation of one-loop 3-point functions} 
\label{se:3point} 

The scalar and tensor 3-point functions are defined in \refapp{app:def}.
We first describe the evaluation of their high-energy expansions and then
list the results.

\subsection{Calculation}
 
In order to calculate the high-energy approximation of the scalar 3-point 
function, we start from the Feynman-parameter representation
\begin{equation}
C_0 (p_1,p_2,m_0,m_1,m_2) = -\int_0^\infty \d x_0 \d x_1 \d x_2 
   \frac { \delta(1-x_0-x_1-x_2) }
         { g(x_0,x_1,x_2)- (p_1-p_2)^2 x_1 x_2 } 
\label{3feynrepr}
\end{equation} 
with 
\begin{equation}
g(x_0,x_1,x_2)=  m_0^2 x_0+m_1^2 x_1+m_2^2 x_2-p_1^2 x_0 x_1
                 -p_2^2 x_0 x_2 - \im \epsilon.
\end{equation}
We search for an asymptotic expansion in the limit
\begin{equation}
r=|(p_1-p_2)^2| \gg m_0^2, m_1^2, m_2^2, p_1^2, p_2^2,
\end{equation}
including all terms of order $1/(p_1-p_2)^2$.
Looking at the specific results derived for the high-energy approximation
of the W-pair production \cite{W-pair}, 
we notice that the scalar 3-point function has
the asymptotic form:
\begin{equation}
C_0(p_1,p_2,m_0,m_1,m_2) = \frac 1 r 
     \left( \frac{c_2} 2 \ln^2 (r) + c_1 \ln (r) +c_0 \right)
   + \O \left( \frac 1 {r^2} \right), \quad r \rightarrow \infty. 
\label{3expansion}
\end{equation}
This asymptotic expansion cannot be extracted by performing a simple 
Taylor expansion or by using the general methods described in
\citere{Ch88}. It is convenient to use the Mellin-transform technique,
because each pole of the Mellin transform is uniquely related 
to one of the terms in the asymptotic expansion \refeq{3expansion}.

If the function $f(r)$ fulfills certain integrability conditions
(see \citere{Bleistein}), the Mellin transform
\begin{equation}
\M [f, \xi ]=\int^\infty _0 r^{\xi -1} f(r) \d r
\end{equation}
converges absolutely and is holomorphic in a vertical strip
$\alpha < \Re (\xi ) < \beta $
of the complex plane.
The corresponding analytical continuation contains poles in the
complex half-planes $\Re (\xi) \ge \beta$ and 
$\Re (\xi) \le \alpha$.
The inversion of the Mellin transform
\begin{equation}
\label{eq:invMellintransform}
f(r) = \frac 1 {2 \pi \im} 
       \int^{c+{\scriptsize \im} \infty }_{c-{\scriptsize \im} \infty } 
       r^{-\xi } 
       \M [f, \xi ] \d \xi 
\quad \mbox{with} \quad  \alpha < c < \beta 
\end{equation}
can be performed by closing the integration contour over the right-hand
complex half-plane at infinity.
According to  residue theorem $f(r)$ is obtained from
the poles of the Mellin transform in the right-hand half-plane
$\Re (\xi )\ge \beta $.

The inverse Mellin transform of a single pole of order $(n+1)$ is given by
\begin{equation}
\label{mellintransformpole}
\frac 1 {2 \pi \im} 
     \int^{c+{\scriptsize \im} \infty }_{c-{\scriptsize \im} \infty }
      \d \xi r^{-\xi} \frac{n!}{(\xi_0-\xi)^{n+1}}
     = \frac 1 {r^{\xi_0}} \ln^n (r),
\end{equation}
\ie a term that appears in the asymptotic expansion \refeq{3expansion}.
{}From this it is obvious that 
we get an asymptotic expansion for $r \rightarrow \infty$
by closing the integration contour in the right-hand complex half-plane. 
Moreover, the leading terms originate from the poles 
of the Mellin transform $\M [f, \xi ]$
on the right edge of the convergence domain,
\ie with $\Re (\xi_0)=\beta$.
Thus, we can  write down the poles of the Mellin transform that determine
the asymptotic expansion \refeq{3expansion} of the
scalar 3-point function 
\begin{equation}
\M[C_0,\xi] = 
     \frac{c_2}{(1-\xi)^3} + \frac{c_1}{(1-\xi)^2} + \frac{c_0}{(1-\xi)}
   + \O ( (1-\xi )^0 )
\label{3mellinexpansion}
\end{equation}
with the same constants $c_i$ as in \refeq{3expansion} 
[see \refeq{mellintransformpole}].

In order to evaluate (multiple) Mellin transforms
of one-loop integrals we use the generalized Feynman-parameter 
representation \cite{Scharf}
\begin{equation}
\int_0^\infty \d r_0 \cdots \d r_n  
\frac{\prod_{i=0}^n r_i^{\xi_i-1} \delta (1- \sum_{i=0}^n \alpha_i r_i)}
     {(\sum_{i=0}^n A_i r_i)^{\xi}}=
     \frac{\prod_{i=0}^n \Gamma(\xi_i)}
          {\Gamma(\xi) \prod_{i=0}^n A_i^{\xi_i}},
\end{equation} 
where $\xi=\sum_{i=0}^n \xi_i$, $\Re (\xi_i )> 0$, $\alpha_i\ge0$, 
and the $A_i$ are not on the negative real axis. 
Choosing  $\alpha_i=0$ for $i>0$ and  $\alpha_0=1$
and performing the integration over $r_0$
we get the following formula for a multiple Mellin transform:
\begin{equation}
\int_0^\infty \d r_1 \cdots \d r_n
\frac{r_1^{\xi_1-1} \cdots r_n^{\xi_n-1}}
     {(A_0+A_1 r_1+ \cdots +A_n r_n)^{\xi}}=
     \frac{\prod_{i=0}^n \Gamma(\xi_i)}
          {\Gamma(\xi) \prod_{i=0}^n A_i^{\xi_i}}
\label{mellintransformation}
\end{equation} 
with $\xi_0=\xi-\sum_{i=1}^n \xi_i$.
This formula can directly be used to evaluate the
Mellin transform of \refeq{3feynrepr} upon identifying:
\begin{equation}
A_0=g(x_0,x_1,x_2), \quad  A_1=(-\sign-\im \epsilon)x_1x_2 
\quad \mbox{and} \quad
r_1=r=|(p_1-p_2)^2| 
\end{equation}
with $\sign=(p_1-p_2)^2/r=\pm1$ and $n=1$. 
In order to ensure the
validity of \refeq{mellintransformation},
an infinitesimal imaginary part of the $ \im \epsilon$-prescription
is given to both $A_0$ and $A_1$.
After integration over $x_0$ we find 
the Mellin transform of the 3-point function,
which is holomorphic in the vertical strip $0<\Re (\xi)<1$,
\begin{equation} 
\label{eq:MT3int}
\M [C_0,\xi ] = -\frac{\Gamma (\xi) \Gamma (1- \xi)} 
                          {(-\sign-\im \epsilon)^{\xi}}
   \int_0^1 \d x_1 \int_0^{1-x_1} \d x_2 
   \frac 1 {h(x_1,x_2)^{1-\xi} x_1^\xi x_2^\xi }
\end{equation}
with $h(x_1,x_2)=g(1-x_1-x_2,x_1,x_2)$.
According to the above discussion, the asymptotic expansion is
determined from the poles at $\xi=1$.
As $\Ga(1-\xi)$ includes a single pole, we need the expansion of the integral 
about $\xi=1$ including the order $\O((1-\xi)^0)$.

First we assume $h(0,0)=m_0^2 \neq 0$.
Naive expansion of $h(x_1,x_2)^{-(1-\xi)}$ about $x_1=0$ and $x_2=0$
would yield two infinite series of pole terms contributing 
to the high-energy approximation namely the Taylor expansions of 
$h(x_1,0)^{-(1-\xi)}$ and $h(0,x_2)^{-(1-\xi)}$.
Therefore, we use
\begin{equation}
\label{eq:expansion}
\frac 1 {h(x_1,x_2)^{1-\xi}} =
    \frac 1 {h(x_1,0)^{1-\xi}}
  + \frac 1 {h(0,x_2)^{1-\xi }}
  - \frac 1 {h(0,0)^{1-\xi}}
  + \O(x_1 x_2) \O(1-\xi),
\end{equation}
such that the terms that are not explicitly written out 
involve the product of $x_1$ and $x_2$,
and get a compact expansion of the Mellin transform:
\begin{eqnarray} 
\nonumber
\M [C_0,\xi ] &=& -\frac{\Gamma (\xi) \Gamma (1- \xi)} 
                             {(-\sign-\im \epsilon)^{\xi}}
      \Bigg\{\frac 1 {h(0,0)^{1-\xi} (1-\xi)^2}-\frac{\pi^2} 6\\
&&{}
             -\int_0^1 \d x \frac{1}{x} \ln \left(\frac{h(x,0)}{h(0,0)}\right)
             -\int_0^1 \d x \frac{1}{x} \ln \left(\frac{h(0,x)}{h(0,0)}\right)
             +\O(1-\xi)
      \Bigg\}.
\end{eqnarray} 
{}From the inversion formula \refeq{eq:invMellintransform}
or by comparing \refeq{3expansion} with \refeq{3mellinexpansion}
we obtain the high-energy approximation of the 3-point function:
\begin{eqnarray} 
\nonumber
&&{} C_0 (p_1,p_2,m_0,m_1,m_2) = \frac 1 {(p_1-p_2)^2}
       \Bigg\{\frac 1 2 
              \ln^2 \left(\frac{-(p_1-p_2)^2- \im \epsilon}{h(0,0)}\right)
\\
&&{}
\hspace*{1cm}
              -\int_0^1 \d x \frac{1}{x} \ln \left(\frac{h(x,0)}{h(0,0)}\right)      
              -\int_0^1 \d x \frac{1}{x} \ln \left(\frac{h(0,x)}{h(0,0)}\right)
       \Bigg\}
     + \O \left( \frac 1 {(p_1-p_2)^4} \right).  
\end{eqnarray} 
\par 

For $h(0,0)=m_0^2=0$ the expansion \refeq{eq:expansion} breaks down.
Therefore, we extend the integration domain of the Mellin transform 
\refeq{eq:MT3int}
to a square ($0 \le x_i \le 1, i=1,2$) and divide this square into two 
sectors\footnote{This 
                 sector decomposition is similar to the one used in the proof
                 of the convergence theorem for Feynman amplitudes 
                 \cite{Hepp}.}:
\begin{equation}
x_1 > x_2 \quad \mbox{and} \quad x_2 > x_1.
\end{equation}
It can easily be seen that the extension has no poles at $\xi=1$,
apart from the one contained in the prefactor $\Ga(1-\xi)$, and 
the integral yields 
\begin{equation}
   \int_0^1 \d x_1 \int_{1-x_1}^1 \d x_2
   \frac 1 {h(x_1,x_2)^{1-\xi} x_1^\xi x_2^\xi }
   = \frac{\pi^2}{6} + \O(1-\xi).
\end{equation}
By substituting $x_2 \rightarrow x_1 \tx_2$ in the first sector
and $x_1 \rightarrow x_2 \tx_1$ in the second sector
we get the following result for the Mellin transform:
\begin{eqnarray} 
\label{eq:MT3int1}
\nonumber
\M [C_0,\xi ] &=&
   -\frac{\Gamma (\xi) \Gamma (1- \xi)} 
                          {(-\sign-\im \epsilon)^{\xi}}
   \Bigg\{  \int_0^1 \d x_1 \int_0^1\d \tx_2 
            \frac 1 {\left(
            h(x_1,x_1 \tx_2)/x_1\right)^{1-\xi} x_1^\xi \tx_2^\xi}\\
&&{}
          + \int_0^1 \d \tx_1 \int_0^1 \d x_2 
            \frac 1 {\left(
            h(x_2 \tx_1,x_2)/x_2\right)^{1-\xi} \tx_1^\xi x_2^\xi}
          - \frac{\pi^2} 6 
          + \O(1-\xi)
   \Bigg\}
\end{eqnarray}
with 
\begin{eqnarray}
\nonumber
 h(x_1,x_1 \tx_2)/x_1&=&     
     m_1^2-p_1^2+(m_2^2-p_2^2) \tx_2
   + (p_1^2 + p_2^2 \tx_2)(1+\tx_2) x_1- \im \epsilon,\\
 h(x_2 \tx_1,x_2)/x_2&=&
     m_2^2-p_2^2+(m_1^2-p_1^2) \tx_1
   + (p_1^2 \tx_1+ p_2^2)(\tx_1+1) x_2- \im \epsilon.
\end{eqnarray}
For $m_1^2 \neq p_1^2$ and $m_2^2 \neq p_2^2$ 
the two integrals in \refeq{eq:MT3int1} can be 
evaluated similarly to \refeq{eq:MT3int} in the case $m_0^2 \neq 0$.

If $m_1^2=p_1^2\neq 0$ we perform a further sector decomposition 
in the first integral of  \refeq{eq:MT3int1}.
The case $m_2^2=p_2^2\neq 0$ is covered by exchanging the indices
$1$ and $2$.
The remaining cases are either IR-singular ($m_0=0$, $m_1^2=p_1^2$,
$m_2^2=p_2^2$) or mass-singular ($m_0^2=m_1^2=p_1^2=0$ or
$m_0^2=m_2^2=p_2^2=0$).
Thus, we can construct high-energy approximations for all
scalar 3-point functions that are neither IR- nor mass-singular.
The singular integrals are also covered by our results
if the singularities are regularized with masses.
Our general results 
were checked by comparing with the specific expressions in \citere{W-pair}.

The high-energy approximation of tensor 3-point integrals can be 
evaluated in the same way:

Consider first the coefficient functions $C_{1 \cdots 1}$.
They have the same Feynman-parameter representation as 
the scalar 3-point function \refeq{3feynrepr}, but with an additional 
factor $x_1^i$ in the numerator [see \refeq{3feyn}] 
which prevents the Mellin transform from being singular at $x_1=0$.
Consequently, the integrand can be expanded with respect to $x_2$,
and terms of order $\O(x_2)$ are not contributing to 
the high-energy behaviour.
As a consequence, the results are independent of $m_2^2$ and $p_2^2$.

The functions $C_{1 \cdots 12 \cdots 2}$ have no singularities at 
$x_1=0$ and $x_2=0$ owing to the numerator factor $x_1^i x_2^j$.
The function $\Ga(1-\xi)$ contributes 
the single pole $1/(1-\xi )$.
Since we can set $\xi=1$ in the integral, 
the result is independent of the internal and external masses.

The remaining coefficient functions of the 2- and 3-point
functions are UV-divergent.
They have no IR or mass singularities,
because the integrand of the Feynman-parameter representation 
has a logarithmic structure.
It can be easily checked
with the help of the Mellin-transform technique that
the high-energy approximation is obtained
by neglecting the internal and external masses.

In contrast to the asymptotic expansion of the 
scalar 3-point function, those of the coefficient functions 
of the tensor 2- and 3-point integrals include no dilogarithms and
need not be split into several different cases. 
All results for these tensor integrals were derived using the Mellin-transform 
technique and checked by tensor-integral reduction.

\subsection{Results}
\label{se:3results}

The following results hold for 3-point functions in the limit
\begin{equation}
|(p_1-p_2)^2| \gg m_0^2, m_1^2, m_2^2, p_1^2, p_2^2,
\end{equation}
and for 2-point functions in the limit
\begin{equation}
|p^2| \gg m_0^2, m_1^2.
\end{equation}

\subsubsection{Scalar 3-point function}
For the scalar 3-point function we have to distinguish several different
cases:\\[1\baselineskip]
$m_0^2\neq 0$:
\begin{eqnarray}
\nonumber 
C_0(p_1,p_2,m_0,m_1,m_2) &\sim & \frac 1 {(p_1-p_2)^2} 
   \Bigg\{  \frac 1 2 \ln^2 
            \left( \frac{-{(p_1-p_2)^2}-\im \epsilon}
                        {m_0^2-\im \epsilon}
            \right)\\ 
&&{} 
\hspace*{2cm}
          + I_C(p_1^2,m_0,m_1) + I_C(p_2^2,m_0,m_2) 
   \Bigg\},
\end{eqnarray}
where
\begin{eqnarray} \label{IC}
\nonumber
I_C(p_1^2,m_0,m_1)&=&- \int_0^1 \d x \frac 1 x \ln 
            \left(  1
                  + \frac{m_1^2-m_0^2-p_1^2}{m_0^2-\im \epsilon}x
                  + \frac{p_1^2}{m_0^2-\im \epsilon}x^2
            \right)\\
&=&
     \sum_\pm 
     \Li_2 \left(\frac{2 p_1^2}
                      {m_0^2-m_1^2+p_1^2
                       \pm \kappa(p_1^2,m_0^2- \im \epsilon, 
                                  m_1^2- \im \epsilon
                                 )}
           \right)
\end{eqnarray}
with $\kappa(a,b,c)=\sqrt{a^2+b^2+c^2-2ab-2ac-2bc}$;\\[1\baselineskip]
$m_0^2=0, m_1^2 \neq p_1^2, m_2^2\neq p_2^2$:
\begin{eqnarray}
C_0(p_1,p_2,m_0,m_1,m_2) &\sim & \frac 1 {(p_1-p_2)^2}
   \Bigg\{  \ln \left(  \frac{-{(p_1-p_2)^2}-\im \epsilon}
                         {m_1^2-p_1^2-\im \epsilon}
            \right)
            \ln \left( \frac{-(p_1-p_2)^2-\im \epsilon}
                        {m_2^2-p_2^2-\im \epsilon} 
            \right)\\
&&{} 
\hspace*{1.8cm}
          + \Li_2 
            \left( -\frac{p_1^2}{m_1^2-p_1^2-\im \epsilon} \right)
          + \Li_2 
            \left( -\frac{p_2^2}{m_2^2-p_2^2-\im \epsilon} \right) 
    \Bigg\};\nn
\end{eqnarray}
$m_0^2 =0, m_1^2=p_1^2 \neq 0, m_2^2\neq p_2^2$:
\begin{eqnarray}
C_0(p_1,p_2,m_0,m_1,m_2) &\sim & \frac 1 {(p_1-p_2)^2}
   \Bigg\{  \ln \left( \frac{-{(p_1-p_2)^2}-\im \epsilon}
                          {m_2^2-p_2^2-\im \epsilon}
            \right) 
            \ln
            \left( \frac{-{(p_1-p_2)^2}-\im \epsilon}
                          {p_1^2-\im \epsilon} 
            \right)\\
&&{} 
\hspace*{.8cm}
          + \frac 1 2  \ln^2 \left( \frac{-{(p_1-p_2)^2}-\im \epsilon}
                          {m_2^2-p_2^2-\im \epsilon}
            \right)
          + \Li_2 \left( -\frac{p_2^2}{m_2^2-p_2^2-\im \epsilon} \right)
          + \frac{\pi^2} 6 \nn
   \Bigg\}.
\end{eqnarray}
The remaining cases are found by exchanging the indices $1$ and $2$
or are IR- or mass-singular. 
Specific examples of these general results can be found in App.~A of
the second paper of \citere{W-pair}.

\subsubsection{Tensor 2- and 3-point functions}
\label{se:3tensor}

The coefficient functions of the  tensor 2- and 3-point integrals,
defined in \refapp{app:def},  read in the
high-energy limit ($k,j>0$): 
\begin{eqnarray}
B_{\cdots} (p,m_0,m_1) &\sim& B_{\cdots} (p,0,0),\\
C_{00\cdots}(p_1^2,(p_1-p_2)^2,p_2^2,m_1,m_0,m_2)
&\sim& 
C_{00\cdots}(0,(p_1-p_2)^2,0,0,0,0),\\
C_{\underbrace{\mbox{\scriptsize 1$\cdots$1}}_k
   \underbrace{\mbox{\scriptsize 2$\cdots$2}}_j
  }(p_1,p_2,m_0,m_1,m_2) 
&\sim& 
      \frac{(-1)^{k+j}(k-1)! (j-1)!}{(p_1-p_2)^2 \; (k+j)!},
\end{eqnarray}
\begin{equation}
C_{\underbrace{\mbox{\scriptsize 1$\cdots$1}}_k} (p_1,p_2,m_0,m_1,m_2) 
\sim 
   \frac 1 {(p_1-p_2)^2} 
   \{  B_{\underbrace{\mbox{\scriptsize 1$\cdots$1}}_{k-1}}(p_1-p_2,0,0)
     - B_{\underbrace{\mbox{\scriptsize 1$\cdots$1}}_{k-1}}(p_1,m_0,m_1)
   \}.
\end{equation}
The most important functions are explicitly given by:
\begin{eqnarray}
B_0(p,0,0)&=&\Delta -\ln \left( \frac{-p^2-\im \epsilon}
                                       {\mu^2 }
                           \right) +2,\\
B_1(p,0,0)&=&-\frac 1 2 B_0(p,0,0),\\
B_{11}(p,0,0)&=&\frac 1 3 B_0(p,0,0)+\frac 1 {18},\\
B_{00}(p,0,0)&=&- \frac{p^2}{12}B_0(p,0,0)
                      - \frac{p^2}{18},
\end{eqnarray}
and
\begin{eqnarray}
C_{12} (p_1,p_2,m_0,m_1,m_2) &\sim & \frac 1 {2 (p_1-p_2)^2} ,\\
C_{112} (p_1,p_2,m_0,m_1,m_2) &\sim & - \frac 1 {6 (p_1-p_2)^2} ,\\
C_{00} (p_1,p_2,m_0,m_1,m_2) &\sim & 
     \frac 1 4 B_0(p_1-p_2,0,0)+\frac 1 4, \\
C_{001} (p_1,p_2,m_0,m_1,m_2) &\sim & 
     -\frac 1 {12}B_0(p_1-p_2,0,0)-\frac 1 {18}.
\end{eqnarray}

\section{High-energy approximations of one-loop 4-point functions}
\label{se:4point} 
The definitions of the scalar and tensor 4-point functions are given in 
\refapp{app:def}.
The high-energy expansion of the scalar 4-point function can also
be evaluated using the Mellin-transform technique. 
Therefore, we sketch only the most
important aspects and, in particular, the differences to the case of the
3-point function discussed above. For the
tensor 4-point functions the Mellin-transform technique breaks down and we
have to rely on tensor-integral reduction. 

\subsection{Calculation}

High-energy approximation of the scalar 4-point 
function means that both $|(k_1+k_2)^2|$ and 
$|(k_1+k_4)^2|$ 
are large compared with the internal and external masses:
\begin{equation}
|(k_1+k_2)^2|, |(k_1+k_4)^2| \gg 
m_0^2, m_1^2, m_2^2, m_3^2, k_1^2, k_2^2, k_3^2, k_4^2.
\end{equation}
Therefore, we need a double Mellin transform in order to evaluate the
expansion.
The scalar 4-point function has the Feynman-parameter representation
\begin{eqnarray}
\nonumber
\hspace*{-1cm}
D_0(k_1,k_1+k_2,k_1+k_2+k_3,m_0,m_1,m_2,m_3)&=&\\
&&{}
\hspace*{-8cm}
   \int^\infty_0 \d x_0 \d x_1 \d x_2 \d x_3
   \frac{\delta (1-x_0 -x_1 -x_2 -x_3 )}
        {\{  g(x_0,x_1,x_2,x_3)
           - (\sign_1+\im \epsilon) r_1 x_0 x_2
           - (\sign_2+\im \epsilon) r_2 x_1 x_3\}^2}
\end{eqnarray}
with 
\begin{equation} 
g(x_0,x_1,x_2,x_3)=
     \sum_{n=0}^3 m_n^2 x_n
   - k_1^2 x_0 x_1 - k_2^2 x_1 x_2 -k_3^2 x_2 x_3-k_4^2 x_3 x_0
   - \im \epsilon ,
\end{equation}
$r_1=|(k_1+k_2)^2|$, $r_2=|(k_1+k_4)^2|$, 
$\sign_1=(k_1+k_2)^2/r_1$, 
$\sign_2=(k_1+k_4)^2/r_2$, where $k_4=-k_1-k_2-k_3$.
As above, we have included an infinitesimal imaginary part 
in the terms involving the large parameters $(k_1+k_2)^2$ and $(k_1+k_4)^2$.
Thus, we can use the generalized Feynman-parameter representation 
\refeq{mellintransformation} in order to perform the double Mellin
transform resulting in:
\begin{eqnarray} \label{MD0}
\nonumber
\M [D_0,\xi_1,\xi_2]&=&
   \frac{\Gamma(\xi_1 ) \Gamma(\xi_2) \Gamma(2-\xi_1-\xi_2)}
        {\left( -\sign_1 - \im \epsilon \right)^{\xi_1}
         \left( -\sign_2 - \im \epsilon \right)^{\xi_2}}
\\
&&{}
   \times \int^\infty_0 \d x_0 \d x_1 \d x_2 \d x_3
   \frac{\delta (1-x_0 -x_1 -x_2 -x_3 )}
        { g(x_0,x_1,x_2,x_3)^{2-\xi_1-\xi_2} 
          x_0^{\xi_1} x_1^{\xi_2} x_2^{\xi_1} x_3^{\xi_2}}.
\end{eqnarray}
{}From the explicit results of \citere{W-pair} we expect the following
form for the high-energy expansion of the scalar 4-point function:
\begin{eqnarray}
D_0 &\sim& \frac 1 {r_1 r_2} 
     \sum_{i,j=0}^2 \frac {c_{ij}} {i! j!} \ln^i (r_1) \ln^j (r_2)
     + \O\left(\frac 1 {r_1^2 r_2}\right)
     + \O\left(\frac 1 {r_1 r_2^2}\right)
     , \quad r_1,r_2 \rightarrow \infty  
\label{4expansion}
\end{eqnarray}
with $c_{22}=c_{12}=c_{21}=0$. 
Accordingly, the Mellin transform should have the following meromorphic 
structure at $\xi_1\approx1$ and $\xi_2\approx1$:
\begin{eqnarray}
\M [D_0,\xi_1,\xi_2]&=&
     \sum_{i,j=0}^2
     \frac {c_{ij}}{(1-\xi_1)^{i+1}(1-\xi_2)^{j+1}} 
     + \O((1-\xi_1)^0)+\O((1-\xi_2)^0).
\label{4mellinexpansion}
\end{eqnarray}
Note that terms that involve only poles in one of the variables $\xi_i$
are irrelevant for the asymptotic expansion, as they do not contribute
to the inverse Mellin transform.
 
In contrast to the scalar 3-point function,
we divide the integration domain of the Mellin transform 
into four sectors and expand each sector separately:
\begin{equation}
x_0 > x_1,x_2,x_3, \quad 
x_1 > x_2,x_3,x_0, \quad 
x_2 > x_3,x_0,x_1, \quad 
x_3 > x_0,x_1,x_2.
\end{equation}
The contribution of each sector can be associated to
one side of the box and must be decomposed into
several cases like for the scalar 3-point function. 
The third Gamma function in \refeq{MD0} involves a pole $1/(2-\xi_1-\xi_2)$.
This pole does not fit into the structure \refeq{4mellinexpansion}.
It is essential for our approach that this pole 
is cancelled in the total result for the Mellin transform, 
although it is present in each single sector. 
  
Because all sectors yield integrals of the same form, 
we consider only the sector $x_0 > x_1,x_2,x_3$. 
We substitute $x_i \rightarrow x_0 x_i$, $i=1,2,3$, in \refeq{MD0}
and obtain for the integral:
\begin{eqnarray}
\Se_0 (\xi_1,\xi_2)&=&   
   \int_0^1 \d x_1 \d x_2 \d x_3 
   \frac 1
        { h(x_1,x_2,x_3)^{2-\xi_1-\xi_2} 
          x_1^{\xi_2} x_2^{\xi_1} x_3^{\xi_2}
        } 
\label{4S0}
\end{eqnarray}
with 
\begin{eqnarray}
\nonumber
\lefteqn{h(x_1,x_2,x_3) =  
     m_0^2 + (m_1^2+m_0^2-k_1^2) x_1
   + (m_2^2+m_0^2) x_2 + (m_3^2+m_0^2-k_4^2) x_3} \qquad \\
&&{}
   + (m_1^2 x_1 + m_2^2 x_2 + m_3^2 x_3)(x_1+x_2+x_3)
   - k_2^2 x_1 x_2 - k_3^2 x_2 x_3 - \im \epsilon.
\end{eqnarray}
   
As for the scalar 3-point function, we consider 
first the case $m_0^2 \neq 0$ and use
\begin{eqnarray}
\nonumber
\frac 1 {h(x_1,x_2,x_3)^{2-\xi_1-\xi_2}}
&=&
  \frac 1 {h(x_1,0,0)^{2-\xi_1-\xi_2}}
+ \frac 1 {h(0,0,x_3)^{2-\xi_1-\xi_2}}
- \frac 1 {h(0,0,0)^{2-\xi_1-\xi_2}}\\
&&{}
+ (2-\xi_1-\xi_2)(\O(x_2)+\O(x_1 x_3)).
\end{eqnarray}
The terms that have not been written down explicitly involve only poles
in one of the variables $\xi_i$ and are hence not relevant for the
asymptotic expansion.
Thus, we find
\begin{eqnarray}
\nonumber
\Se_0 (\xi_1,\xi_2) &\sim & 
     \frac {(2-\xi_1-\xi_2)}
           {(1-\xi_1)(1-\xi_2)}
     \Bigg\{\frac 1 {h(0,0,0)^{1-\xi_2}(1-\xi_2)^2}\\ 
\nonumber
&&{}
            - \int_0^1 \d x \frac 1 x 
              \ln \left(\frac {h(x,0,0)}{h(0,0,0)}\right)
            - \int_0^1 \d x \frac 1 x 
              \ln \left(\frac {h(0,0,x)}{h(0,0,0)}\right)
     \Bigg\}\\
&&
     - \frac 1 {(1-\xi_2)^3}
     + (2-\xi_1-\xi_2) 
       \{\O((1-\xi_1)^0)+\O((1-\xi_2)^0)\}.           
\label{4sectorentwicklung}
\end{eqnarray}
The factor $(2-\xi_1-\xi_2)$ in the first and third terms cancels the pole 
contained in $\Ga(2-\xi_1-\xi_2)$ in \refeq{MD0}. If we combine all four
sectors this pole is also cancelled in the contributions of the second
term:
\begin{eqnarray}
\nonumber
\sum_{n=0}^3 \Se_n (\xi_1,\xi_2) &=&
       - \frac 2 {(1-\xi_1)^3}-\frac 2 {(1-\xi_2)^3}+\O(2-\xi_1-\xi_2)\\
\nonumber
&=& - \frac{2(2-\xi_1-\xi_2)}{(1-\xi_1)(1-\xi_2)}
         \left\{  \frac 1 {(1-\xi_1)^2} 
                + \frac 1 {(1-\xi_2)^2}
                - \frac 1 {(1-\xi_1)(1-\xi_2)}
         \right\}\\
&&{}
       + \O(2-\xi_1-\xi_2).
\label{4sectoren}
\end{eqnarray}
This holds as well in the other cases with $m_0^2=0$ considered below.

It is then straightforward to construct the expansion of the Mellin transform
and evaluate the high-energy approximation by comparing \refeq{4expansion} 
and \refeq{4mellinexpansion}.
Because the calculation is lengthy but analogous to
the one for the scalar 3-point function, we only make some remarks.
The integrals involving the logarithms in \refeq{4sectorentwicklung}
lead to contributions of the form
\begin{eqnarray} \label{ID}
I_D(k_1^2,m_0,m_1)&=&- \int_0^1 \d x \frac 1 x \ln
            \left(  1
                  + \frac{m_1^2+m_0^2-k_1^2}{m_0^2-\im \epsilon}x
                  + \frac{m_1^2}{m_0^2-\im \epsilon}x^2
            \right).
\end{eqnarray}
Each of these integrals can be associated with one corner of the box and
for each corner two integrals appear in the sum of all sectors.
Using the relation 
\begin{equation}
              I_D(k_1^2,m_0,m_1)+I_D(k_1^2,m_1,m_0) = 
              I_C(k_1^2,m_0,m_1)+I_C(k_1^2,m_1,m_0)-\frac {\pi^2} 3
\end{equation}
with $m_0^2, m_1^2 \neq 0$,
all these integrals can be expressed by those appearing in the
asymptotic expansion of the 3-point functions. Finally the asymptotic
expansion of the scalar 4-point function can be written in terms of the
asymptotic expansions of the four corresponding scalar 3-point functions and a
universal term.

Next we consider the case $m_0^2=0$. 
We decompose the integration domain of the integral $\Se_0$ into three sectors
\begin{equation}
x_1 > x_2, x_3, \quad
x_2 > x_3, x_1, \quad
x_3 > x_1, x_2,
\end{equation}
and arrive at
\begin{eqnarray}
\label{eq:S0split}
\nonumber
     \Se_0
&=&
     \int_0^1 \d x_1 \d x_2 \d x_3 
     \frac 1 {h_1(x_1,x_2,x_3)^{2-\xi_1-\xi_2} 
     x_1^{\xi_2} x_2^{\xi_1} x_3^{\xi_2}}\\
\nonumber
&&{}
  + \int_0^1 \d x_1 \d x_2 \d x_3 
    \frac 1 {h_2(x_1,x_2,x_3)^{2-\xi_1-\xi_2} 
    x_1^{\xi_2} x_2^{\xi_2} x_3^{\xi_2}}\\
&&{}
  + \int_0^1 \d x_1 \d x_2 \d x_3 
    \frac 1 {h_3(x_1,x_2,x_3)^{2-\xi_1-\xi_2} 
    x_1^{\xi_2} x_2^{\xi_1} x_3^{\xi_2}}
\end{eqnarray}
with the polynomials
\begin{eqnarray} 
\nonumber
h_1(x_1,x_2,x_3)&=& h(x_1,x_1 x_2,x_1 x_3)/x_1,\\
\nonumber
h_2(x_1,x_2,x_3)&=& h(x_2 x_1,x_2,x_2 x_3)/x_2,\\
h_3(x_1,x_2,x_3)&=& h(x_3 x_1,x_3 x_2,x_3)/x_3.
\end{eqnarray}

If $m_1^2 \neq k_1^2$ and $m_3^2 \neq k_4^2$, the
first and third integral are analogous to \refeq{4S0}.
The second integral can be evaluated by
performing an expansion of the integrand with respect to $(2-\xi_1-\xi_2)$,
\begin{equation}
\label{eq:integral2}
\int_0^1 \d x_1 \d x_2 \d x_3 
     \frac 1 {h_2(x_1,x_2,x_3)^{2-\xi_1-\xi_2} 
              x_1^{\xi_2} x_2^{\xi_2} x_3^{\xi_2}}
=  \frac 1 {(1-\xi_2)^3}
   + (2-\xi_1-\xi_2) \O((1-\xi_1)^0),
\end{equation}
which is true no matter whether $h_2(0,0,0) = 0$ or not.
Notice that \refeq{4sectoren} holds also
after performing sector decomposition, because the pole $1/(1-\xi_2)^3$
of the second integral in \refeq{eq:S0split} [see
\refeq{eq:integral2}] cancels 
one of the poles in the first and third integral 
[see \refeq{4sectorentwicklung}].

If $m_1^2=k_1^2\neq 0$ or $m_3^2=k_4^2\neq 0$ 
we obviously have to apply a further sector 
decomposition to the first or third integral, respectively.
In this way, all 4-point functions which contain neither 
IR nor mass singularities can be approximated 
for high energies. Again the singular ones are covered by our approach
once they are regularized with masses.
It turns out, that in all cases the scalar 4-point function can be
expressed in a universal way in terms of the four corresponding 3-point
functions.

In contrast to the case of the scalar 4-point function,
the pole $1/(2-\xi_1-\xi_2)$ does not cancel 
in the Mellin transform of the coefficient functions of the tensor 4-point 
integrals.
Because the poles of the variable $\xi_1$ and $\xi_2$ are coupled
in the Mellin transform, the resulting asymptotic expansion depends on the
order of the inversion of the Mellin transforms with  respect to $\xi_1$
and $\xi_2$. 
Integrating first over $\xi_1$ and then over $\xi_2$ results in an
approximation for $r_1 \gg r_2 \gg m_i^2, k_i^2$, integrating first over
$\xi_2$ and then over $\xi_1$ yields an approximation for $r_2 \gg r_1
\gg m_i^2, k_i^2$. While these approximations coincide for the scalar
4-point function, where the poles in $\xi_1$ and $\xi_2$ factorize, 
they differ for the tensor 4-point functions.
Therefore, we have evaluated the approximations for the tensor 4-point
integrals by reducing them to scalar integrals with 
FeynCalc \cite{FeynCalc} and determining the high-energy limit 
by using the approximations for the scalar integrals 
with the help of Mathematica \cite{Mathematica}. 
As can be seen from the results in \refse{se:4tensor}, 
the pole $1/(2-\xi_1-\xi_2)$ in the Mellin transform corresponds to a 
factor $1/(\si_1r_1+\si_2r_2)$ ($1/u$ in the notation of \refse{se:4tensor}) 
in the high-energy approximation of the integrals.

\subsection{Results}
 
The results are valid for 
\begin{equation}
|s|, |t|, |u| \gg k_1^2, k_2^2, k_3^2, k_4^2, m_0^2, m_1^2, m_2^2, m_3^2
\end{equation}
with 
$s=(k_1+k_4)^2$, $t=(k_1+k_2)^2$ and $u=(k_1+k_3)^2$.
All 3-point functions in this section
are implicitly understood in the high-energy limit.

\subsubsection{Scalar 4-point function}

The high-energy approximation of the scalar 4-point function
can be reduced to the  approximations of the four corresponding 3-point
functions as follows:
\begin{eqnarray}
\nonumber 
&&{}
D_0(k_1,k_1+k_2,k_1+k_2+k_3,m_0,m_1,m_2,m_3) \sim -\frac 1 {st}
\left[\ln^2\left(\frac{-t-\im \epsilon}{-s- \im \epsilon}\right)
+\pi^2 \right]
\\
\nonumber 
&&{}
\hspace*{2cm}
+ \frac 1 s C_0(k_1,-k_2,m_1,m_0,m_2) 
 +  \frac 1 t C_0(k_2,-k_3,m_2,m_1,m_3)\\
&&{}
\hspace*{2cm}
+ \frac 1 s C_0(k_3,-k_4,m_3,m_2,m_0) 
 +  \frac 1 t C_0(k_4,-k_1,m_0,m_3,m_1).
\end{eqnarray}
Note that this result holds for all non-singular cases, in particular
for arbitrary masses.
The necessary split-up into different cases has to
be done only at the level of 3-point functions.
Specific examples of this general result can be found in App.~A of
the second paper of \citere{W-pair}.

\subsubsection{\label{se:4tensor}Tensor 4-point functions}
The  high-energy approximations of the
coefficient functions of the tensor 4-point integrals,
defined in \refapp{app:def}, can be written as:
\begin{eqnarray}
D_1 &\sim & - \frac 1 {2su} (\Ln^2+\pi^2) -\frac 1 s C_0,\\
D_{11} &\sim & - 
    \frac t {2su^2} (\Ln^2+\pi^2) 
  - \frac 1 {su}   
    \Ln
  + \frac 1 s (C_0+C_1+C_2),\\
D_{12} &\sim & -
     \frac 1 {2u^2} (\Ln^2+\pi^2) 
  + \frac 1 {su}  
    \Ln
  - \frac 1 s C_2,\\
D_{13} &\sim &  
    \frac 1 {2u^2} (\Ln^2+\pi^2) 
  - \frac 1 {su}  
    \Ln,\\
\nonumber 
D_{111} &\sim & - 
     \frac{t^2}{2su^3} (\Ln^2+\pi^2) 
   - \frac{3t+s}{2su^2}  
     \Ln
   + \frac{t+2s}{2tsu}\\
&&{}
   - \frac 1 s (C_0+C_1+C_2-C_{11}-C_{22}),\\
D_{112} &\sim & -
     \frac t {2u^3} (\Ln^2+\pi^2) 
   + \frac{t-s}{2su^2}  
     \Ln
   - \frac 1 {2tu}
   - \frac 1 s C_{22},\\
D_{113} &\sim & 
     \frac t {2u^3} (\Ln^2+\pi^2) 
   - \frac{t-s}{2su^2}  
     \Ln
   - \frac 1 {2su},\\
D_{123} &\sim &  
     \frac{s-t}{4u^3} (\Ln^2+\pi^2) 
   - \frac 1 {u^2}  
     \Ln
   + \frac 1 {2su},\\
\nonumber 
D_{1111} &\sim & -
     \frac{t^3}{2su^4} (\Ln^2+\pi^2) 
   - \frac{11t^2+7ts+2s^2}{6su^3}  
     \Ln
   + \frac{2t^2+7ts+4s^2}{2tsu^2}\\
&&{}
   + \frac 1 s (C_0+C_1+C_2-C_{11}-C_{22}+C_{111}+C_{222}),\\
D_{1112} &\sim & -
     \frac{t^2}{2u^4} (\Ln^2+\pi^2) 
   + \frac{2t^2-5ts-s^2}{6su^3}  
     \Ln
   - \frac{2t+s}{2tu^2}
   - \frac 1 s C_{222},\\
D_{1113} &\sim & 
     \frac{t^2}{2u^4} (\Ln^2+\pi^2) 
   - \frac{2t^2-5ts-s^2}{6su^3}  
     \Ln
   - \frac t {2su^2},\\
\nonumber 
D_{1122} &\sim & -
     \frac{ts}{2u^4} (\Ln^2+\pi^2) 
   + \frac{t^2+5ts-2s^2}{6su^3}  
     \Ln
   - \frac{t^2-ts+s^2}{6tsu^2}\\
&&{}
   + \frac 1 s (C_{22}+C_{222}),\\
D_{1123} &\sim & 
     \frac{2ts-t^2}{6u^4} (\Ln^2+\pi^2) 
   - \frac{5t-s}{6u^3}  
     \Ln
   + \frac{t-2s}{6su^2},\\
D_{1133} &\sim & 
     \frac{2t^2-ts}{6u^4} (\Ln^2+\pi^2) 
   - \frac{t^2-5ts}{6su^3}  
     \Ln
   - \frac{2t-s}{6su^2},\\
D_{1223} &\sim & 
     \frac{s^2-2ts}{6u^4} (\Ln^2+\pi^2) 
   + \frac{t-5s}{6u^3}  
     \Ln
   + \frac{t+4s}{6su^2},\\
D_{00} &\sim & -
     \frac 1 {4u} (\Ln^2+\pi^2) ,\\
D_{001} &\sim & - \frac t {8u^2} (\Ln^2+\pi^2) 
   - \frac 1 {4u}  
     \Ln,\\
D_{0011} &\sim & - 
     \frac{t^2}{12u^3} (\Ln^2+\pi^2) 
   - \frac{3t+s}{12u^2}  
     \Ln
   - \frac 1 {12u},\\
D_{0012} &\sim & -
     \frac{ts}{12u^3} (\Ln^2+\pi^2) 
   + \frac{t-s}{12u^2}  
     \Ln
   + \frac 1 {12u},\\
D_{0013} &\sim & 
     \frac{ts-t^2}{24u^3} (\Ln^2+\pi^2) 
   - \frac t {6u^2}  
     \Ln
   - \frac 1 {12u},\\
D_{0000} &\sim & -
     \frac{ts}{48u^2} (\Ln^2+\pi^2) 
   - \frac s {24u}  
     \Ln
   + \frac 5 {72}
   + \frac 1 {24}B_0(t,0,0),
\end{eqnarray}
where
\begin{equation}
\Ln = \ln \left(\frac{-t-\im \epsilon}{-s-\im \epsilon}\right).
\end{equation}
The arguments of the 3- and 4-point functions are given by:
\begin{eqnarray}
\nonumber 
D_{\cdots}&=&D_{\cdots}(k_1,k_1+k_2,k_1+k_2+k_3,m_0,m_1,m_2,m_3),\\
C_{\cdots}&=&C_{\cdots}(k_1,-k_2,m_1,m_0,m_2).
\end{eqnarray}
For the 3-point integrals the approximations given in 
\refse{se:3results} have to be inserted. 
Note that only one kind of 3-point functions appears.
Moreover, the coefficient functions with indices  $00$ or
$1$ and $3$ are independent of all
internal and external masses and do not involve dilogarithms.

\section{Conclusion}

We have calculated high-energy approximations 
of the basic one-loop integrals 
necess\-ary for scattering processes of two on-shell particles into two
on-shell particles ($2\to2$ processes).
All kinematical variables have been assumed to be large compared with the 
internal and external masses, and all terms with the leading power of the
energy have been kept. 

We used the Mellin-transform technique and overcame the difficulties
in expanding the Mellin transform by sector decomposition.
As a consequence, the results for the scalar integrals fall 
into several different cases.
It turned out that the high-energy approximation
of the scalar 4-point functions is just given by the four corresponding
3-point functions and an universal term.
The tensor 2- and 3-point integrals were 
expanded with the help of the Mellin-transform technique and checked by using
tensor-integral reduction.
The approximations for these
integrals are, in contrast to the one for the scalar 3-point function,
very short and include no dilogarithms.
The tensor 4-point integrals were first reduced to 
scalar functions and then approximated resulting in compact expressions
in terms of 3-point functions and logarithms.

We have listed
high-energy approximations for the complete set of scalar one-loop integrals 
that appears in $2\to2$ processes.
In addition, we
have provided explicit results for the corresponding tensor integrals.
In the case of the 4-point function we have restricted ourselves to
tensor integrals with at most four Lorentz indices.
This set is sufficient for calculations in the 't Hooft--Feynman gauge. 
The tensor integrals with
more indices that appear in more general gauges can be easily reduced to
those given here using tensor-integral reduction.

With the help of our results and tensor-integral reduction
any one-loop matrix element for $2\to2$ scattering processes 
can be approximated for energies 
much larger than the internal and external masses.
Moreover, our approximations are applicable to the decay of a very
heavy particle into two light ones when all internal masses are also light. 
Since our results contain only the leading terms they allow to extract
the leading contributions to matrix elements in the high-energy limit,
\ie the non-vanishing terms for $2\to2$ matrix elements.
 
\begin{appendix}

\section{Appendix}
\subsection{Definition of one-loop integrals and tensor-coefficient functions}
\label{app:def}

We use dimensional regularization denoting the space--time dimension by $D$.
The one-loop functions are defined by:
\begin{eqnarray}
A_0 (m_0)&=&\frac{(2\pi \mu)^{4-D}}{\im \pi^2} \int \d^D q \frac{1}{D_0}\\
B_{\{ 0,\mu,\mu \nu \} }(p_1,m_0,m_1)&=&
     \frac{(2 \pi \mu)^{4-D}}{\im \pi^2}
     \int \d^D q \frac{\{ 1,q_\mu ,q_\mu q_\nu\} }{D_0 D_1},\\
\hspace*{-0.5cm}
C_{\{ 0,\mu ,\mu \nu ,\mu \nu \rho \} }(p_1,p_2,m_0,m_1,m_2)&=&
     \frac{(2\pi \mu)^{4-D}}{\im \pi^2} 
     \int \d^D q \frac{\{1,q_\mu ,q_\mu q_\nu ,q_\mu q_\nu q_\rho \} }
                     {D_0 D_1 D_2},\\
\nonumber
\hspace*{-0.5cm}
D_{\{ 0,\mu ,\mu \nu ,\mu \nu \rho ,\mu \nu \rho \sigma \} } 
      (p_1,p_2,p_3,m_0,m_1,m_2,m_3) =
\hspace*{-2.5cm}
&&{}
\\
&=& \frac{(2\pi \mu)^{4-D}}{\im \pi^2} 
    \int \d^D q 
    \frac{\{ 1,q_\mu ,q_\mu q_\nu ,q_\mu q_\nu q_\rho ,
             q_\mu q_\nu q_\rho q_\sigma 
         \}}
    {D_0 D_1 D_2 D_3}
\end{eqnarray}
with $D_0=q^2-m_0^2+\im \epsilon$,
     $D_i=(q+p_i)^2-m_i^2+\im \epsilon$,
     $i\ge1$.
The tensor integrals can be decomposed into Lorentz tensors 
constructed of (linearly independent) external momenta $p_{i\mu}$ and the 
metric tensor $g_{\mu \nu}$ and tensor-coefficient functions as follows:
\begin{eqnarray}
B_\mu &=&  B_1 p_{1 \mu},\\[1ex]
B_{\mu \nu} &=& 
    B_{11} p_{1 \mu} p_{1 \nu}
  + B_{00} g_{\mu \nu} ,\\
C_\mu &=& \sum _{i=1}^2 C_i p_{i \mu},\\
C_{\mu \nu} &=& 
    \sum _{i,j=1}^2 C_{ij} p_{i \mu} p_{j \nu}
  + C_{00} g_{\mu \nu} ,\\ 
C_{\mu \nu \rho} &=& 
    \sum _{i,j,k=1}^2 C_{ijk} p_{i \mu} p_{j \nu} p_{k \rho} 
  + \sum _{i=1}^2 C_{00i} 
  \{  g_{\mu \nu} p_{i \rho}
    + g_{\nu \rho} p_{i \mu}
    + g_{\mu \rho} p_{i \nu } 
  \},\\
D_\mu &=& \sum _{i=1}^3 D_i p_{i \mu},\\
D_{\mu \nu} &=& 
    \sum _{i,j=1}^3 C_{ij} p_{i \mu} p_{j \nu}
  + D_{00} g_{\mu \nu} ,\\
D_{\mu \nu \rho} &=& 
    \sum _{i,j,k=1}^3 D_{ijk} p_{i \mu} p_{j \nu} p_{k \rho}
  + \sum _{i=1}^3 D_{00i} 
  \{  g_{\mu \nu} p_{i \rho}
    + g_{\nu \rho} p_{i \mu}
    + g_{\mu \rho} p_{i \nu } 
  \},\\
\nonumber 
D_{\mu \nu \rho \sigma} &=&
    \sum _{i,j,k,l=1}^3 D_{ijkl} 
    p_{i \mu} p_{j \nu} p_{k \rho} p_{l \sigma}
  + D_{0000} 
  \{ g_{\mu \nu} g_{\rho \sigma}
     + g_{\mu \rho} g_{\nu \sigma} 
     + g_{\mu \sigma} g_{\nu \rho} 
  \}\\
\nonumber 
&&{}
  + \sum _{i,j=1}^3 D_{00ij} 
    \{  g_{\mu \nu} p_{i \rho} p_{j \sigma}
      + g_{\mu \rho} p_{i \nu} p_{j \sigma}
      + g_{\mu \sigma} p_{i \nu} p_{j \rho}\\  
&&{}
\hspace*{2.1cm}
      + g_{\nu \rho} p_{i \mu} p_{j \sigma}
      + g_{\nu \si} p_{i \mu} p_{j \rho} 
      + g_{\rho \sigma} p_{i \mu} p_{j \nu} 
    \}.
\end{eqnarray}

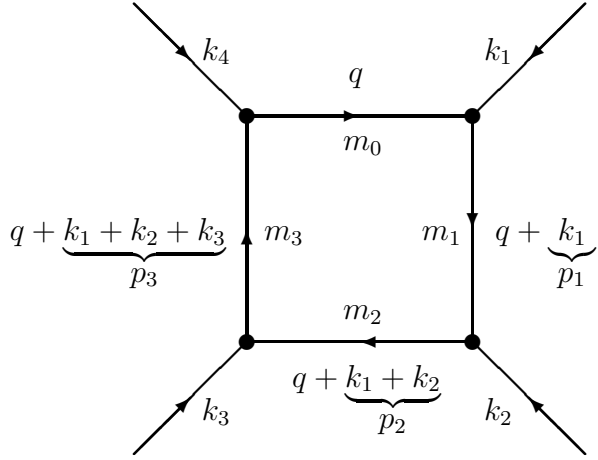
\begin{figure}
\begin{center}
\setlength{\unitlength}{1.5cm}
\begin{picture}(4,4)
\thicklines
\put(1,1){\circle*{0.15}}
\put(3,1){\circle*{0.15}}
\put(1,3){\circle*{0.15}}
\put(3,3){\circle*{0.15}}
\put(0,0){\vector(1,1){0.5}}
\put(4,4){\vector(-1,-1){0.5}}
\put(0,4){\vector(1,-1){0.5}}
\put(4,0){\vector(-1,1){0.5}}
\put(1,1){\line(-1,-1){1}}
\put(3,3){\line(1,1){1}}
\put(1,3){\line(-1,1){1}}
\put(3,1){\line(1,-1){1}}
\put(1,3){\line(1,0){2}}
\put(3,3){\line(0,-1){2}}
\put(3,1){\line(-1,0){2}}
\put(1,1){\line(0,1){2}}
\put(1,3){\vector(1,0){1}}
\put(3,3){\vector(0,-1){1}}
\put(3,1){\vector(-1,0){1}}
\put(1,1){\vector(0,1){1}}
\put(1.85,2.7){\makebox(0,0)[lb]{$m_0$}}
\put(2.55,1.9){\makebox(0,0)[lb]{$m_1$}}
\put(1.85,1.2){\makebox(0,0)[lb]{$m_2$}}
\put(1.15,1.9){\makebox(0,0)[lb]{$m_3$}}
\put(0.6,3.5){\makebox(0,0)[lb]{$k_4$}}
\put(3.1,3.5){\makebox(0,0)[lb]{$k_1$}}
\put(0.6,0.3){\makebox(0,0)[lb]{$k_3$}}
\put(3.1,0.3){\makebox(0,0)[lb]{$k_2$}}
\put(1.9,3.3){\makebox(0,0)[lb]{$q$}}
\put(3.2,1.9){\makebox(0,0)[lb]{$q
   + \underbrace{k_1}_{\mbox{$p_1$}}$}}
\put(1.4,0.6){\makebox(0,0)[lb]{$q
   + \underbrace{k_1+k_2}_{\mbox{$p_2$}}$}}
\put(-1.1,1.9){\makebox(0,0)[lb]{$q
   + \underbrace{k_1+k_2+k_3}_{\mbox{$p_3$}}$}}
\end{picture}
\caption{Conventions for 4-point integrals}
\label{fi:d0}
\end{center}
\end{figure}

Since the tensor-coefficient functions are scalars and depend only on 
kinematical invariants, internal and external masses, 
we can write their arguments as: 
\begin{eqnarray}
&&{} C_{\cdots} (p_1,p_2,m_0,m_1,m_2)=
C_{\cdots} (p_1^2,(p_2-p_1)^2,p_2^2,m_1,m_0,m_2),\\
\nonumber
&&{} D_{\cdots} (k_1,k_1+k_2,k_1+k_2+k_3,m_0,m_1,m_2,m_3)=
\\
&&{}
\hspace*{2cm}
=D_{\cdots} (k_1^2,k_2^2,k_3^2,k_4^2,(k_1+k_2)^2,(k_1+k_4)^2,
            m_0,m_1,m_2,m_3),
\end{eqnarray}
where for convenience we introduce the momenta $k_i$ of the external particles
(\cf \reffi{fi:d0}). 

\subsection{Feynman-parameter representation of one-loop functions}

The UV divergences of the following integrals are contained in the quantity
\begin{equation}
\Delta = \frac 2 {4-D} - \gamma_E + \ln(4 \pi ),
\end{equation}
where  $\gamma_E$ is Euler's constant.

\subsubsection{1- and 2-point functions}

The tensor-coefficient functions of the 1- and 2-point integrals can be 
represented as ($i\ge0$):
\begin{eqnarray}
A_0(m_0)&=& m_0^2 
   \left\{  \Delta 
          - \ln \left(\frac{m_0^2-\im \epsilon}{\mu^2}\right)
          + 1 
   \right\},\\
B_{\underbrace{\mbox{\scriptsize 1$\cdots$1}}_i}(p_1,m_0,m_1)&=& (-1)^i
   \left\{  \frac 1 {i+1} \Delta
          - \int_0^\infty \d x_0 \d x_1 x_1^i \delta (1-x_0-x_1)
            \ln \left(\frac{M_B^2}{\mu^2} \right)
   \right\}, \qquad
\label{2feyn}
\\
\nonumber 
\hspace*{-0.5cm}
B_{00 \underbrace{\mbox{\scriptsize 1$\cdots$1}}_i}(p_1,m_0,m_1)&=&    
   (-1)^i \frac 1 2
   \Bigg\{  \left(  \frac{m_0^2}{(i+1)(i+2)} 
                  + \frac{m_1^2}{(i+2)} 
                  - \frac{p_1^2}{(i+2)(i+3)}
            \right) (\Delta + 1)
\\
&&{}
\hspace*{1.5cm}       
          - \int_0^\infty \d x_0 \d x_1 x_1^i 
            \delta (1- x_0 -x_1) 
            M_B^2 \ln \left(\frac{M_B^2}{\mu^2} \right)
   \Bigg\}
\end{eqnarray}
with $M_B^2= m_0^2 x_0 + m_1^2 x_1 - p_1^2 x_0 x_1 - \im \epsilon$.
Note that the scalar 2-point function $B_0$ is given by \refeq{2feyn} with
$i=0$.

\subsubsection{\label{se:3feyn}3-point functions}

The tensor-coefficient functions of the 3-point integrals  
$C_{\cdots} (p_1,p_2,m_0,m_1,m_2)$
allow for the following Feynman-parameter representations \mbox{($i,j\ge0$)}:
\begin{eqnarray}
C_{\underbrace{\mbox{\scriptsize 1$\cdots$ 1}}_{i} 
   \underbrace{\mbox{\scriptsize 2$\cdots$ 2}}_{j}
  }&=&
    - (-1)^{i+j} \int_0^\infty \d x_0 \d x_1 \d x_2 
      \frac{x_1^i x_2^j \delta (1- x_0 - x_1 - x_2)}{M_C^2},
\label{3feyn}
\\
\nonumber
C_{00\underbrace{\mbox{\scriptsize 1$\cdots$ 1}}_{i} 
   \underbrace{\mbox{\scriptsize 2$\cdots$ 2}}_{j}
  }&=&
    (-1)^{i+j} \frac 1 2
    \Bigg\{  \frac{i!j!}{(i+j+2)!} \Delta           
\\
&&{}
\hspace*{1cm}
           - \int_0^\infty \d x_0 \d x_1 \d x_2 x_1^i x_2^j 
             \delta (1- x_0 - x_1 - x_2) 
             \ln\left(\frac{M_C^2}{\mu^2}\right)
    \Bigg\} \qquad
\end{eqnarray}
with 
$M_C^2 = m_0^2 x_0 + m_1^2 x_1 + m_2^2 x_2 - p_1^2 x_0 x_1 - p_2^2 x_0 x_2 
          - (p_1-p_2)^2 x_1 x_2 - \im \epsilon $.
The scalar 3-point function is given by \refeq{3feyn} with $i=j=0$.

\subsubsection{4-point functions}

The Feynman-parameter representations for the tensor-coefficient
functions of the \mbox{4-point} integrals 
$D_{\cdots} (k_1,k_1+k_2,{k_1+k_2+k_3},\allowbreak m_0,m_1,m_2,m_3)$
read ($i,j,k\ge0$):
\begin{eqnarray}
D_{\underbrace{\mbox{\scriptsize 1$\cdots$ 1}}_i 
   \underbrace{\mbox{\scriptsize 2$\cdots$ 2}}_j 
   \underbrace{\mbox{\scriptsize 3$\cdots$ 3}}_k
  } &=&
     (-1)^{i+j+k} 
     \int_0^\infty \d x_0 \d x_1 \d x_2 \d x_3 
     \frac{x_1^i x_2^j x_3^k \delta (1- \sum_{n=0}^3 x_n)}{M_D^4},
\label{4feyn}
\\
D_{00
   \underbrace{\mbox{\scriptsize 1$\cdots$ 1}}_i 
   \underbrace{\mbox{\scriptsize 2$\cdots$ 2}}_j 
   \underbrace{\mbox{\scriptsize 3$\cdots$ 3}}_k
  } &=&
     - \frac 1 2 (-1)^{i+j+k}
       \int_0^\infty \d x_0 \d x_1 \d x_2 \d x_3
       \frac{x_1^i x_2^j x_3^k \delta (1- \sum_{n=0}^3 x_n)}{M_D^2},\qquad \\
\nonumber
D_{0000
   \underbrace{\mbox{\scriptsize 1$\cdots$ 1}}_i 
   \underbrace{\mbox{\scriptsize 2$\cdots$ 2}}_j 
   \underbrace{\mbox{\scriptsize 3$\cdots$ 3}}_k
  } &=&
     \frac 1 4 (-1)^{i+j+k} 
     \Bigg\{  \frac{i!j!k!}{(i+j+k+3)!} \Delta \\
&&{}
            - \int_0^\infty \d x_0 \d x_1 \d x_2 \d x_3 x_1^i x_2^j x_3^k 
              \delta (1- \mbox{$\sum_{n=0}^3$} x_n) 
              \ln\left(\frac{M_D^2}{\mu^2}\right)
     \Bigg\}
\end{eqnarray}
with $M_D^2=  \sum\limits_{n=0}^3 m_n^2 x_n 
            - k_1^2 x_0 x_1 - k_2^2 x_1 x_2 - k_3^2 x_2 x_3 - k_4^2 x_3 x_0
            - (k_1+k_2)^2 x_0 x_2 - (k_1+k_4)^2 x_1 x_3 - \im \epsilon$
and $k_4=-k_1-k_2-k_3$.
The scalar 4-point function is given by \refeq{4feyn} with $i=j=k=0$.

\end{appendix}

\section*{Acknowledgement}
We thank S.~Dittmaier for useful discussions and a careful reading
of the manuscript.

\def\refname{REFERENCES}


\begin{thebibliography}{99}
\itemsep 2pt plus 2pt minus 1pt
\frenchspacing
\newcommand{\anp}[3]{{\sl Ann.~Phys.} {\bf #1} (19#2) #3}
\newcommand{\app}[3]{{\sl Acta~Phys.~Pol.} {\bf #1} (19#2) #3}
\newcommand{\cmp}[3]{{\sl Commun. Math. Phys.} {\bf #1} (19#2) #3}
\newcommand{\cpc}[3]{{\sl Comp. Phys. Commun.} {\bf #1} (19#2) #3}
\newcommand{\fp}[3]{{\sl Fortschr. Phys.} {\bf #1} (19#2) #3}
\newcommand{\ijmp}[3]{{\sl Int. J. Mod. Phys.} {\bf #1} (19#2) #3}
\newcommand{\jetp}[3]{{\sl JETP} {\bf #1} (19#2) #3}
\newcommand{\jetpl}[3]{{\sl JETP Lett.} {\bf #1} (19#2) #3}
\newcommand{\jmp}[3]{{\sl J. Math. Phys.} {\bf #1} (19#2) #3}
\newcommand{\mpl}[3]{{\sl Mod. Phys. Lett.} {\bf #1} (19#2) #3}
\newcommand{\nc}[3]{{\sl Nuovo Cimento} {\bf #1} (19#2) #3}
\newcommand{\nim}[3]{{\sl Nucl. Instr. Meth.} {\bf #1} (19#2) #3}
\newcommand{\np}[3]{{\sl Nucl. Phys.} {\bf #1} (19#2)~#3}
\newcommand{\npb}{{\sl Nucl. Phys.} {\bf B} }
\newcommand{\pl}[3]{{\sl Phys. Lett.} {\bf #1} (19#2) #3}
\newcommand{\pr}[3]{{\sl Phys. Rev.} {\bf #1} (19#2) #3}
\newcommand{\prl}[3]{{\sl Phys. Rev. Lett.} {\bf #1} (19#2) #3}
\newcommand{\ptp}[3]{{\sl Prog. Theor. Phys.} {\bf #1} (19#2) #3}
\newcommand{\sptp}[3]{{\sl Prog. Theor. Phys. Suppl.} {\bf #1} (19#2) #3}
\newcommand{\sjnp}[3]{{\sl Sov. J. Nucl. Phys.} {\bf #1} (19#2) #3}
\newcommand{\tmp}[3]{{\sl Theor. Math. Phys.} {\bf #1} (19#2) #3}
\newcommand{\zp}[3]{{\sl Z. Phys.} {\bf #1} (19#2) #3}
\newcommand{\vj}[4]{{\sl #1~}{\bf #2} (19#3) #4}
\newcommand{\ej}[3]{{\bf #1} (19#2) #3}
\newcommand{\vjs}[2]{{\sl #1~}{\bf #2}}

\bibitem{LEP95}
D.~Bardin, W.~Hollik and G.~Passarino (eds.), {\it Reports of the
Working Group on Precision Calculations for the Z Resonance\/}
(CERN-95-03, Geneva, 1995).

\bibitem{W-pair} 
   W.~Beenakker, A.~Denner, S.~Dittmaier and  R.~Mertig,  
   \pl{B317}{93}{622};\\
   W.~Beenakker, A.~Denner, S.~Dittmaier, R.~Mertig and  T.~Sack,  
   \np{B410}{93}{245}.

\bibitem{FeynCalc} 
   R. Mertig, M. B\"ohm and A. Denner, 
   \cpc{64}{91}{345}.
 
\bibitem{Pa79}
L.M. Brown and R.P.~Feynman, \pr{85}{52}{231};\\
G.~Passarino and M.~Veltman, \np{B160}{79}{151}.

\bibitem{Me65} D.B.~Melrose, \nc{XL~A}{65}{181}.

\bibitem{tH79} 
G.~'t Hooft and M.~Veltman, \np{B153}{79}{365};\\
W.~Beenakker and A.~Denner, \np{B338}{90}{349};\\
A.~Denner, U.~Nierste and R.~Scharf, \np{B367}{91}{637}.

\bibitem{Ch88} K.G.~Chetyrkin, \tmp{75}{88}{346} and
\ej{76}{88}{809}; MPI Munich preprint MPI-PAE/PTh 13/91;\\
S.G.~Gorishny, \np{B319}{89}{633};\\
V.A.~Smirnov, \cmp{134}{90}{109};\\
for a recent review see V.A. Smirnov, \mpl{A10}{95}{1485}.

\bibitem{Sm93} V.A.~Smirnov, \pl{B309}{93}{397}.

\bibitem{Bleistein} 
   N.~Bleistein and R.A.~Handelsman, 
   {\sl Asymptotic Expansions of Integrals}, 
   (Holt, Rinehart and Winston, New York, 1975).

\bibitem{Scharf} 
     R.~Scharf and J.B.~Tausk, \np{B412}{94}{523}.

\bibitem{Hepp}
    K. Hepp,
    \cmp{2}{66}{301}.

\bibitem{Mathematica} 
   S.~Wolfram,  
   {\sl Mathematica---A System for Doing Mathematics by Computer}
   (Addison-Wesley, New York, 1988).

\end{thebibliography}
\end{document}